\documentclass{ws-ijmpa}
\usepackage[super,compress]{cite}

\usepackage{amssymb}
\usepackage{amsmath}
\usepackage{epsfig}
\usepackage[dvips]{color}
\usepackage{graphicx}

\newcommand{\be}{\begin{equation}}
\newcommand{\ee}{\end{equation}}
\newcommand{\bea}{\begin{eqnarray}}
\newcommand{\eea}{\end{eqnarray}}
\newcommand{\beas}{\begin{eqnarray*}}
\newcommand{\eeas}{\end{eqnarray*}}

\newcommand{\nn}{\nonumber}

\begin{document}
\markboth{A. Ayala,J.D. Casta\~no, J.J. Cobos-Mart\1nez, S. Hern\'andez-Ortiz, A.J. Mizher, A. Raya}{Chiral Symm. transition in the LSMq: Counting effective QCD dof from low to high $T$}

%
\catchline{}{}{}{}{}
%

\title{Chiral Symmetry transition in the Linear Sigma Model with quarks: Counting effective QCD degrees of freedom from low to high temperature
}
\author{Alejandro Ayala$^{1,3}$, Jorge David Casta\~no-Yepes$^1$, J.J. Cobos-Mart\'inez$^{2,4}$, Sa\'ul Hern\'andez-Ortiz$^2$, Ana Julia Mizher$^1$, Alfredo Raya$^2$}

\address{
  $^1$Instituto de Ciencias
  Nucleares, Universidad Nacional Aut\'onoma de M\'exico, Apartado
  Postal 70-543, M\'exico Distrito Federal 04510,
  Mexico.\\
  $^2$Instituto de F\1sica y Matem\'aticas, Universidad Michoacana de 
  San Nicol\'as de Hidalgo, Edificio C-3, Ciudad Universitaria, Morelia,
  Michoac\'an 58040, Mexico.\\
  $^3$Centre for Theoretical and Mathematical Physics, and Department of Physics,University of Cape Town, Rondebosch 7700, South Africa\\
  $^4$Laborat\'orio de F\1sica Te\'orica e Computacional - LFTC, Universidade Cruzeiro do Sul, 01506-000, S\~ao Paulo, Brazil}
  
\maketitle

\begin{history}
\received{Day Month Year}
\revised{Day Month Year}
\end{history}

\begin{abstract}
We use the linear sigma model coupled to quarks,
together with a plausible location of the  critical end point (CEP), to study the chiral symmetry transition in the QCD phase diagram. We compute the effective potential at finite temperature and density up to the contribution of the ring diagrams, both in the low and high temperature limits, and use it to compute the pressure and the position of the CEP. In the high temperature regime, by comparing to results from extrapolated lattice data, we determine the model coupling constants. Demanding that the CEP remains in the same location when described in the high temperature limit,  we determine again the couplings and the pressure for the low temperature regime. We show that this procedure gives an average description of the lattice QCD results for the pressure and that the change from the low to the high temperature domains in this quantity can be attributed to the change in the coupling constants which in turn we link to the change in the effective 
degrees of freedom.

\end{abstract}

\ccode{25.75.Nq, 11.30.Rd, 11.15.Tk}

\section{Introduction}\label{I}

In the study of QCD thermodynamics one of the principal goals is to gather accurate knowledge of the phase diagram in the quark chemical potential ($\mu$) versus temperature ($T$) plane, describing the degrees of freedom of strongly interacting matter. Data from the BNL Relativistic Heavy Ion Collider (RHIC)~\cite{RHIC} and the CERN Large Hadron Collider (LHC)~\cite{LHC1,LHC2} show that in heavy-ion collisions a deconfined phase, the so-called Quark Gluon Plasma (QGP), is produced. For vanishing $\mu$, this phase takes place above a (pseudo)critical temperature $T_c$ that lattice QCD calculations have shown to represent a region where an analytic crossover takes place~\cite{Aoki}. The most recent value for this temperature provided by lattice QCD calculations is $T_c=155(1)(8)$ MeV~\cite{HotQCD2} considering 2+1 quark flavors. 

On the other hand for vanishing $T$, a number of different model approaches indicate that the transition along the quark chemical potential axis is strongly first order~\cite{first-order}. Since the first order line originating at $T=0$ cannot end at the $\mu=0$ axis, which corresponds to the starting point of the cross-over line, it must terminate somewhere in the middle of the phase diagram. This point is generally referred to as the critical end point (CEP). Mathematical extensions of lattice calculations, for instance, the Taylor expansion technique~\cite{PRD71} or the Fourier expansion of the grand canonical partition function~\cite{NP153} (which considers an imaginary chemical potential) place the CEP in the region $(\mu^{\mbox{\tiny{CEP}}}/T_c,T^{\mbox{\tiny{CEP}}}/T_c)\sim(1.0-1.4,0.9-0.95)$~\cite{Sharma}. For recent reviews see Refs.~\cite{Ding,AyalaRev}.

The extension of lattice QCD calculations to $\mu\neq 0$ is hindered by the {\it sign problem}~\cite{Forcrand}. Although some mathematical extensions of lattice calculations~\cite{mathlattice} as well as Schwinger-Dyson equation techniques~\cite{Roberts} can be employed in the finite $\mu$ region, the use of effective QCD models continues to be a useful tool to explore a large portion of the phase diagram~\cite{NJL,pNJL,Chqm,pChqm,Ayala1,Ayala2}. As emphasized in Refs.~\cite{Ayala1,Ayala2}, for theories where massless bosons appear, the proper treatment of the plasma screening effects in the calculation of the effective finite temperature potential is paramount to determining the CEP location. The importance of accounting for screening in plasmas was pointed out since the pioneering work in Ref.~\cite{Jackiw} and implemented also in the context of the Standard Model to study the electroweak phase transition~\cite{Carrington}. 

In this work we use the linear sigma model coupled to quarks (LSMq) as an effective model for the strong interactions to determinate the transition lines and the CEP location in the phase diagram. The same model has been previously used to incorporate magnetic field effects on the couplings to explore the influence of the latter on the inverse magnetic catalysis phenomenon~\cite{AyalaIMC}. We compute the effective potential at finite temperature and density in the low and high temperature limits. To account for the plasma screening effects, the computation of the effective potential is carried out up to the contribution of ring diagrams. We use the low temperature expansion to determine the model coupling constants requiring that the CEP location agrees with the one provided by extrapolated lattice results and then compute the pressure. Then, by requiring that the CEP remains in the same location when described from the high temperature behavior of the effective potential, we determine the values of the couplings in that limit and also compute the pressure. We show that the pressure thus computed provides an average description of lattice results and that its change from the low to the high temperature regimes can be attributed to the change in the coupling constants, which in turn arises from the change in the effective degrees of 
freedom from the low to the high temperature regimes.

This paper is organized as follows: In Sec.~\ref{II} we outline the basics of the LSMq. In Sec.~\ref{III} we compute the finite $T$ and $\mu$ effective potential up to the ring diagrams order. The calculation requires knowledge of the self-energy which we also find both in the low and high temperature approximations. In Sec.~\ref{IV} we use the effective potentials found in the high and low-temperature limits to explore the phase diagram and in particular to locate the CEP in the region found by  mathematical extensions of lattice QCD~\cite{Sharma}. The guiding benchmark is to obtain the same CEP location when working in the high and low-temperature approximations. We test our findings by computing the pressure and comparing to lattice QCD results. Finally we summarize and conclude in Sec.~\ref{concl}.

\section{The Linear Sigma Model coupled to Quarks}\label{II}

We start from the LSMq. The Lagrangian density is given by 
\bea
\mathcal{L}&=&\frac{1}{2}(\partial_\mu\sigma)^2+\frac{1}{2}(\partial_\mu\vec{\pi})^2+\frac{a^2}{2}(\sigma^2+\vec{\pi}^2)-\frac{\lambda}{4}(\sigma^2+\vec{\pi}^2)^2\nn\\
&+&i\bar{\psi}\gamma^\mu\partial_\mu\psi-g\bar{\psi}(\sigma+i\gamma_5\vec{\tau}\cdot\vec{\pi})\psi,
\label{Lsigmamodel}
\eea
where $\psi$ is an SU(2) isospin doublet, $\vec{\pi}$ is an isospin triplet and $\sigma$ is an isospin singlet. The neutral pion, $\pi^0$, is taken as the third component of $\vec{\pi}$ and the charged pions as $\pi_{\pm}=\left(\pi_1\mp i\pi_2\right)/2$. We require that the squared mass parameter $a^2$ and the coupling constants $\lambda$ and $g$ are positive.

The spontaneous breaking of symmetry is obtained when the $\sigma$ field develops a vacuum expectation value $v$ that can later be taken as the order parameter of the theory. Thus we shift $\sigma$ as
\bea
\sigma\rightarrow\sigma +v\;,
\eea
so that the Lagrangian density becomes
\bea
\mathcal{L}&=&-\frac{1}{2}\sigma\partial_\mu\partial^\mu\sigma-\frac{1}{2}(3\lambda v^2-a^2)\sigma^2
-\frac{1}{2}\vec{\pi}\partial_\mu\partial^\mu\vec{\pi}
+\frac{a^2}{2}v^2
\nn\\&-&\frac{1}{2}(\lambda  v^2-a^2)\vec{\pi}^2
-\frac{\lambda}{4}v^4+i\bar{\psi}\gamma^\mu\partial_\mu\psi-gv\bar{\psi}\psi+\mathcal{L}_I^b+\mathcal{L}_I^f
\label{aftersymmetrybreaking},
\eea
where
\bea
\mathcal{L}_I^b&=&-\frac{\lambda}{4}\left[(\sigma^2+\pi_0^2)^2+4\pi^+\pi^-(\sigma^2+\pi_0^2+\pi^+\pi^-)\right],\nn\\
\mathcal{L}_I^f&=&-g\bar{\psi}(\sigma+i\gamma_5\vec{\tau}\cdot\vec{\pi})\psi,
\label{interactions}
\eea
describe the interactions among the fields after symmetry breaking. Equation~(\ref{aftersymmetrybreaking}) gives the masses of fields in terms of the order parameter $v$, the mass parameter $a$ and the coupling constants $\lambda$ and 
$g$, namely,
\bea
m_\sigma^2&=&3\lambda v^2-a^2,\nn\\
m_\pi^2&=&\lambda v^2-a^2,\nn\\
m_f&=&gv.
\label{masas}
\eea

We now proceed to use this theory to compute the effective potential at finite temperature and density.

\section{Effective Potential and Self-Energy}\label{III}

In this section we compute the $T$- and $\mu$-dependent effective potential up to ring diagrams order to account for the plasma screening effects. The quark chemical potential is introduced assuming the conservation of the baryon number $\mathcal{Q}$, so that in equilibrium  the system is described by a grand canonical partition function $Z={\mbox{Tr}}\left[\exp\{-({\mathcal{H}}-\mu{\mathcal{Q}})\beta)\}\right]$, with $\beta=1/T$. Using the imaginary-time formalism of finite temperature field theory, this amounts to replace the Matsubara fermion frequencies $i\widetilde{\omega}_n$ by $i\widetilde{\omega}_n-\mu$ when computing the fermion contribution to the effective potential~\cite{LeBellac}. We obtain the effective potential both in the high as well as in the low temperature limits. Since the ring contribution requires calculation of the boson self-energy, we also show the results for this quantity in these regimes. 
\begin{figure}[t]
\centering
\includegraphics[width=0.7\textwidth]{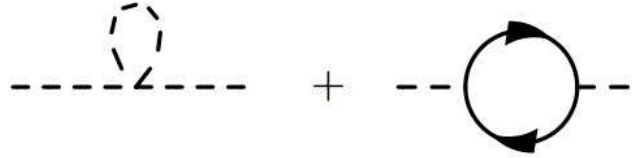}
\caption{Feynman diagrams depicting the contributions to the one-loop boson self-energy. Dashed lines represent bosons whereas solid lines represent fermions. The boson 1-loop diagram receives contributions from the pions and sigma.}
\label{fig1}
\end{figure}

\subsection{High Temperature Approximation}

The effective potential and the self-energy at finite temperature and chemical potential up to the contribution of the ring diagrams in the limit where the masses are small compared to temperature, after mass renormalization at the scale $\widetilde{\mu}=e^{-1/2}a$ have been computed in detail in Ref.~\cite{Ayala1} and are given by
\bea
V^{({\mbox{\small{eff}}})}&=&-\frac{a^2}{2}v^2 + \frac{\lambda}{4}v^4
+\sum_{i=\sigma,\vec{\pi}}\left\{\frac{m_i^4}{64\pi^2}
\left[ \ln\left(\frac{(4\pi T)^2}{2a^2}\right)
-2\gamma_E +1\right]-\frac{\pi^2T^4}{90}\right. \nn\\
&&\left.+ \frac{m_i^2T^2}{24}
-\frac{T}{12 \pi}(m_i^2 + \Pi)^{3/2} \right\} 
-\frac{N_{c}}{16\pi^2}\sum_{f=u,d}
\left\{m_f^4\left[\ln\left(\frac{(4\pi T)^2}{2a^2}\right)\right.\right.\nn\\
&+&1+\left.\psi^0\left(\frac{1}{2}+\frac{i\mu}{2\pi T}\right) +
\psi^0\left(\frac{1}{2}-\frac{i\mu}{2\pi T}\right) \right] \nn\\
&+&8\ m_f^2T^2\left[ \text{Li}_2(-e^{\frac{\mu}{T}}) + \text{Li}_2(-e^{-\frac{\mu}{T}})\right]\nn\\
&-& 
\left. 
32\ T^4 \left[ \text{Li}_4(-e^{\frac{\mu}{T}}) + \text{Li}_4(-e^{-\frac{\mu}{T}}) \right]\right\},
\label{Veff-mid}
\eea
and
\bea
\Pi&=&\frac{\lambda T^2}{2} - \frac{N_fN_cg^2T^2}{\pi^2}
\left[ \text{Li}_2(-e^\frac{\mu}{T}) + \text{Li}_2(-e^{-\frac{\mu}{T}})\right],
\label{self-energy} 
\eea 
respectively, where $\gamma_E\simeq 0.5772$ is the Euler-Mascheroni constant, $\psi_0(z)$ is the digamma function and $\text{Li}_n(x)$ is a polylogarithm function of order $n$. $N_f=2$ and $N_c=3$ are the number of light flavors and colors, respectively. The first (second) term in Eq.~(\ref{self-energy}) is the high temperature 1-loop boson (fermion) contribution. The second term is computed also in the approximation  where the external momentum can be neglected compared to the temperature.

\subsection{Low Temperature Approximation}

Since for the low temperature calculation of the effective potential we need to resort to numerical integration, we start from the original expressions that provide the boson and fermion contributions to the one-loop effective potential 
\bea
V_b^{(1)}&=&T\sum_{i=\sigma,\vec{\pi}}\sum_{n=-\infty}^{\infty}\int\frac{d^3k}{(2\pi)^3}\ln D^{-\frac{1}{2}},\nonumber\\
V_f^{(1)}&=&T\sum_{i=u,d}\sum_{n=-\infty}^{\infty}\int\frac{d^3k}{(2\pi)^3}\ln S,
\label{Vbf}
\eea
where the thermal boson and fermion propagators are given by
\bea
D&=&\frac{1}{k^2+m_i^2+\omega_n^2},\qquad
S\ = \ \frac{1}{k^2+m_i^2+(\widetilde{\omega}_n - i\mu)^2},
\label{DS}
\eea
respectively, with
\bea
\omega_n&=&2n\pi T,\qquad 
\widetilde{\omega}_n \ =\ (2n+1)\pi T,
\label{omegas}
\eea
being the Matsubara frequencies for bosons and fermions, respectively.

\begin{figure*}[t]
\centering
\includegraphics[width=\textwidth]{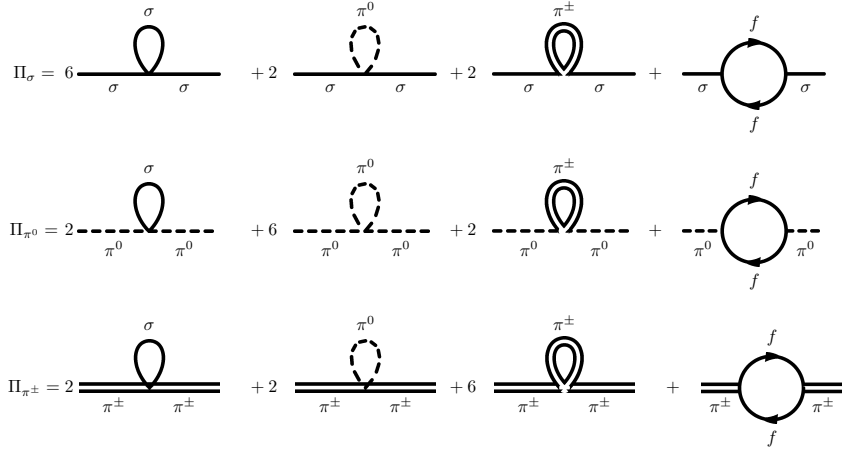}
\caption{Self-energy Feynman diagrams for each boson species together with the corresponding symmetry factors that multiply each boson loop diagram. Notice that since there is no interaction that distinguishes  between charged and neutral pions, the self energy expressions for one and the other species become equal.}
\label{fig2}
\end{figure*}

To account for the plasma screening effects, the boson contribution to the effective potential is computed up to ring diagrams order~\cite{LeBellac}. This contribution is written as
\bea
V_b^{({\mbox{\tiny{ring}}})}=\frac{T}{2}\sum_{i=\sigma,\vec{\pi}}\sum_{n=-\infty}^{\infty}\int\frac{d^3k}{(2\pi)^3}\ln\left[ 1 + \Pi D\right]\;.
\label{Vbring}
\eea
Using the expression for the boson propagator in Eq.~(\ref{DS}), we obtain
\bea
V_b^{({\mbox{\tiny{ring}}})}&=&\frac{T}{2}\sum_{i=\sigma,\vec{\pi}}\sum_{n=-\infty}^{\infty}\int\frac{d^3k}{(2\pi)^3}
\ln\left[ \frac{k^2+m_i^2+\omega_n^2 + \Pi}{k^2+m_i^2+\omega_n^2} \right]\nonumber\\
&=&\frac{T}{2}\sum_{i=\sigma,\vec{\pi}}\sum_{n=-\infty}^{\infty}\int\frac{d^3k}{(2\pi)^3}
\ln\left[k^2+m_i^2+\omega_n^2 + \Pi\right]\nonumber\\
&-&\frac{T}{2}\sum_{i=\sigma,\vec{\pi}}\sum_{n=-\infty}^{\infty}\int\frac{d^3k}{(2\pi)^3}
\ln\left[ k^2+m_i^2+\omega_n^2 \right].
\label{Vbringexpl}
\eea
By adding the 1-loop and the ring diagram boson contribution to the effective potential we obtain
\bea
V_b& \equiv &V_b^{(1)}+V_b^{({\mbox{\tiny{ring}}})}\nonumber\\
& = &\frac{T}{2}\sum_{i=\sigma,\vec{\pi}}\sum_{n=-\infty}^{\infty}
\int\frac{d^3k}{(2\pi)^3}
\ln\left[k^2+m_i^2+\omega_n^2 + \Pi\right]\nonumber\\
&=&T\sum_{i=\sigma,\vec{\pi}}\sum_{n=-\infty}^{\infty}
\int\frac{d^3k}{(2\pi)^3}
\ln \left(D^{({\mbox{\tiny{ring}}})}\right)^{-\frac{1}{2}}.
\label{Vmod}
\eea
Therefore, the calculation, after considering the sum of the 1-loop and ring diagram contributions, is carried out with a boson propagator where $m_i^2\rightarrow m_i^2 + \Pi$.

Let us first look at the boson contribution. In order to work with Eq.~(\ref{Vmod}) for a single boson species with mass $m_i$, we can rewrite the expression as 
\bea
V_{bi}&=&T\sum_{n=-\infty}^{\infty}\int\frac{d^3k}{(2\pi)^3}\int dm_i^2\frac{\partial}{\partial m_i^2}
\ln \left(D^{({\mbox{\tiny{ring}}})}\right)^{-\frac{1}{2}}\nn\\
&=&T\sum_{n=-\infty}^{\infty}\int\frac{d^3k}{(2\pi)^3}\int dm_i^2
\left(-\frac{1}{2}
\left(D^{({\mbox{\tiny{ring}}})}\right)^{-1}\frac{\partial \left(D^{({\mbox{\tiny{ring}}})}\right)}{\partial m_i^2}\right)\nn\\
&=&\frac{1}{2}\int\frac{d^3k}{(2\pi)^3}\int dm_i^2\;T\sum_{n=-\infty}^{\infty}\frac{1}{\omega_n^2+\omega_i^2},
\label{Vb2}
\eea
where
\bea
\omega_i^2\equiv k^2+m_i^2 + \Pi.
\eea
Using the fact that the sum over Matsubara frequencies in Eq.~(\ref{Vb2}) can be written as
\bea
T\sum_{n=-\infty}^{\infty}\frac{1}{\omega_n^2+\omega^2}=\frac{1}{2\omega}\left[1+\frac{2}{e^\frac{\omega}{T}-1}\right],
\eea
we obtain 
\bea
V_b=\sum_{i=\sigma,\vec{\pi}}\int\frac{d^3k}{(2\pi)^3}\left[\frac{\omega_i}{2}+T\ln\left(1-e^{-\frac{\omega_i}{T}}\right)\right]\label{Vb3}
\eea
In order to proceed, we need to compute the self-energy $\Pi$ in the low temperature limit. The diagrams contributing to $\Pi$ are depicted in Fig.~\ref{fig1}.

The self energy $\Pi_i$ for a single boson $i$ is given by 

\bea
\Pi_i=\sum_{j=\sigma,\vec{\pi}}\frac{s_j}{4\pi^2}\int \frac{k^2}{\omega_i}n(\omega_i)dk+\Pi_f\;,
\label{selfenergyboson}
\eea 
where $s_j$ is the symmetry factor that corresponds to each boson loop,
\bea
n(\omega_i)=\frac{1}{\exp\left(\frac{\omega_i}{T}\right)-1},
\label{be}
\eea
and $\Pi_f$ represents the 1-loop fermion contribution to the boson's 
self-energy. Note that for the temperatures of interest, namely, close to the phase transition, $T$ can still be considered large compared to the fermion mass. Therefore, even though the temperature cannot be taken as small when compared to the mass parameter $a$ in the boson sector, it is still a good approximation to consider the large temperature expansion with respect to the mass in the fermion sector. Thus, we take for $\Pi_f$ the same expression as the one in the second term of Eq.~(\ref{self-energy}), namely
\bea
\Pi_f=\frac{N_fN_cg^2T^2}{\pi^2}\left[ \text{Li}_2(-e^{\frac{\mu}{T}}) + \text{Li}_2(-e^{-\frac{\mu}{T}})\right].
\label{Pif}
\eea 

The diagrams contributing to each boson species' self-energy are depicted in Fig.~\ref{fig2} together with its corresponding symmetry factor that can be read off from the interaction Lagrangian.

The explicit expressions for the boson self-energies are given by
\bea
\Pi_b^\sigma&=&\frac{\lambda}{2}\left[3I\left(\sqrt{m_\sigma^2+\Pi_b^\sigma}\right)+2I\left(\sqrt{m_{\pi^{\pm}}^2+\Pi_b^{\pi^{\pm}}}\right)\right.\nn\\
&+&\left.I\left(\sqrt{m_{\pi^{0}}^2+\Pi_b^{\pi^{0}}}\right)\right] + \Pi_f,\nn\\
\Pi_b^{\pi^{0}}&=&\frac{\lambda}{2}\left[I\left(\sqrt{m_\sigma^2+\Pi_b^\sigma}\right)+2I\left(\sqrt{m_{\pi^{\pm}}^2+\Pi_b^{\pi^{\pm}}}\right)\right.\nn\\
&+&\left.3I\left(\sqrt{m_{\pi^{0}}^2+\Pi_b^{\pi^{0}}}\right)\right] + \Pi_f, \nn\\
\Pi_b^{\pi^{\pm}}&=&\frac{\lambda}{2}\left[I\left(\sqrt{m_\sigma^2+\Pi_b^\sigma}\right)+4I\left(\sqrt{m_{\pi^{\pm}}^2+\Pi_b^{\pi^{\pm}}}\right)\right.\nn\\
&+&\left.I\left(\sqrt{m_{\pi^{0}}^2+\Pi_b^{\pi^{0}}}\right)\right] + \Pi_f,
\label{piesFull}
\eea
where
\bea
I(x)&=&\frac{1}{4\pi^2}\int \frac{k^2}{\sqrt{k^2+x^2}}n\left(\sqrt{k^2+x^2}\right)dk.
\label{Idef}
\eea
Given that in our scheme charged and neutral pion masses are equal as a consequence of isospin symmetry, we cannot distinguish between the $\pi^0$ and $\pi^{\pm}$ self-energies. Therefore, Eqs.~(\ref{piesFull}) reduce to the simpler system
\bea
\Pi_b^\sigma&=&\frac{\lambda}{2}\left[3I\left(\sqrt{m_\sigma^2+\Pi_b^\sigma}\right)+3I\left(\sqrt{m_{\pi}^2+\Pi_b^{\pi}}\right)\right]+\Pi_f,\nn\\
\Pi_b^{\pi}&=&\frac{\lambda}{2}\left[I\left(\sqrt{m_\sigma^2+\Pi_b^\sigma}\right)+5I\left(\sqrt{m_{\pi}^2+\Pi_b^{\pi}}\right)\right]+\Pi_f.
\label{pies}
\eea

Note that Eqs.~(\ref{pies}) represent a system of coupled equations for the self energies.  Since the boson masses depend on the order parameter $v$, the solutions will also depend on $v$.

Finally, using Eqs.~(\ref{Vb3}), (\ref{Idef}) and~(\ref{pies}), and after mass renormalization at the scale $\widetilde{\mu}=e^{-1/2}a$, the effective potential in the low temperature approximation is 
\bea
V^{({\mbox{\small{eff}}})}&=&-\frac{a^2}{2}v^2+\frac{\lambda}{4}v^4\nn\\
&+&\sum_{i=\sigma,\vec{\pi}}\left\lbrace\frac{\left(m_i^2+\Pi_i\right)^2}{64\pi^2}\left[\ln\left(\frac{m_i^2+\Pi_i}{4\pi a^2}\right)+\gamma_E-\frac{1}{2}\right]\right.\nn\\
&+&\frac{T}{2\pi^2}\int dk k^2
\left.\ln\left[1-\exp\left(-\frac{\sqrt{k^2+m_i^2+\Pi_i}}{T}\right)\right]\right\rbrace\nn\\
&-& \frac{N_{c}}{16\pi^2}\sum_{f=u,d}
\left\{m_f^4\left[\ln\left(\frac{(4\pi T)^2}{2a^2}\right)+1\right.\right.\nn\\
&+&\left.\psi^0\left(\frac{1}{2}+\frac{i\mu}{2\pi T}\right) +
\psi^0\left(\frac{1}{2}-\frac{i\mu}{2\pi T}\right) \right] \nn\\
&+&8\ m_f^2T^2\left[ \text{Li}_2(-e^{\frac{\mu}{T}}) + \text{Li}_2(-e^{-\frac{\mu}{T}})\right]\nn\\
&-& \left. 32\ T^4 \left[ \text{Li}_4(-e^{\frac{\mu}{T}}) + \text{Li}_4(-e^{-\frac{\mu}{T}}) \right]\right\}.
\label{Vfinal}
\eea

We now proceed to use Eqs.~(\ref{Veff-mid}) and~(\ref{Vfinal}) to study the phase diagram from the low and the high temperature approaches and in particular to locate a CEP. 
In order to determine the model parameters in the low temperature regime,
we use as input the CEP location found in the high temperature approximation in a previous
work~\cite{Karsch}, which lies within the region found by lattice inspired 
calculations~\cite{Sharma}. Our guiding principle then is to find the same CEP location when 
computed in both the low and high temperature approaches. Subsequently we compute the pressure in 
both regimes and compare the results to those of lattice QCD.

\section{Locating the CEP and computing the pressure}\label{IV}

In order to determine the phase boundaries we compute from the effective potential $V^{({\mbox{\small{eff}}})}$ the values of $\mu_c$ and $T_c$ for which $v_0$, the value of the order parameter that minimizes $V^{({\mbox{\small{eff}}})}$, changes from $v_0=0$ to a finite value. For low values of $\mu$ such change is continuous and the corresponding transitions are associated to cross-over transitions in the general case with nonzero current mass but described as second order phase transitions in this approach with zero current mass. Increasing $\mu$, one reaches a pair of values $\mu^{\mbox{\tiny{CEP}}}$ and $T^{\mbox{\tiny{CEP}}}$ for which the change in $v_0$ starts becoming discontinuous. These changes are associated to first order phase transitions. This procedure requires as starting point the fixing of the model parameters, a procedure we explain below.

In order to find the values of  $\lambda$, $g$ and $a$ appropriate for the description of the phase transition, we note that when considering the thermal effects the boson masses are modified since they acquire a thermal component. For $\mu=0$ they are
\bea
m_\sigma^2(T)&=&3\lambda v^2 -a^2 + \frac{\lambda T^2}{2}+\frac{N_fN_cg^2T^2}{6}\nn\\
m_\pi^2(T)&=&\lambda v^2 -a^2 + \frac{\lambda T^2}{2}+\frac{N_fN_cg^2T^2}{6}.
\label{massmod} 
\eea 
\begin{figure}[t]
\centering
\includegraphics[width=0.7\textwidth]{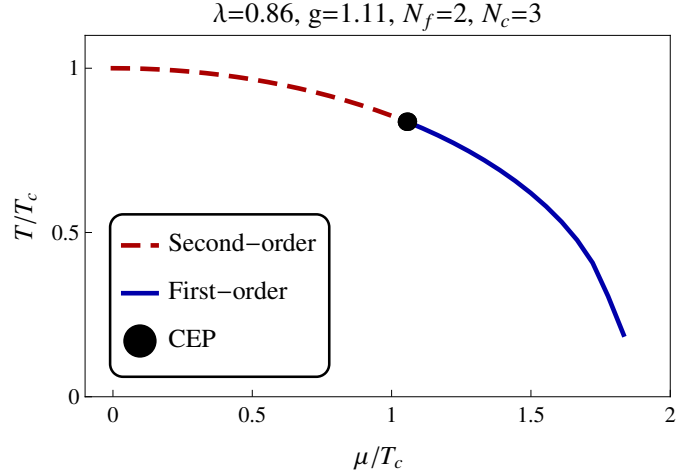}

\caption{(Color on-line) Phase diagram computed from the high-temperature approximation.  This procedure gives as a possible solution $\lambda=0.86$ and $g=1.11$.}
\label{fig5}
\end{figure}

At the phase transition, the curvature of the
effective potential vanishes for $v=0$. Since the boson thermal
masses are proportional to this curvature, they also vanish at
$v=0$. From any of the Eqs.~(\ref{massmod}), we obtain a relation
between the model parameters at $T_c$
\bea
a=T_c\sqrt{\frac{\lambda}{2}+\frac{N_fN_cg^2}{6}}.
\label{relation}
\eea
Furthermore, we can fix the value of $a$ by noting from Eqs.~(\ref{masas}) that the vacuum boson masses satisfy
\bea
a=\sqrt{\frac{m_\sigma^2 - 3m_\pi^2}{2}}.
\label{massvac}
\eea

Since in our scheme we consider two flavors of quarks in the chiral limit, we take $T_c\simeq 170$ MeV~\cite{example} which is slightly larger than $T_c$ obtained in $N_f=2+1$ lattice simulations. From Eqs.~(\ref{relation}) and~(\ref{massvac}) the coupling constants are proportional to $m_\sigma$, given that this is large compared to the pion mass. 

To explore the phase diagram within the high-$T$ approximation, we impose that the couplings $g$ and $\lambda$ are restricted by Eq.~(\ref{relation}). Also, in order to allow for a crossover phase transition for $\mu=0$ with $g,\lambda \sim {\cal O}(1),$ we need that $g^2>\lambda$. A solution consistent with the above requirements gives $\lambda=0.86$ and $g=1.11$. Figure~\ref{fig5} shows the phase diagram and the CEP thus found.

\begin{figure}[t]
\centering
\includegraphics[width=0.7\textwidth]{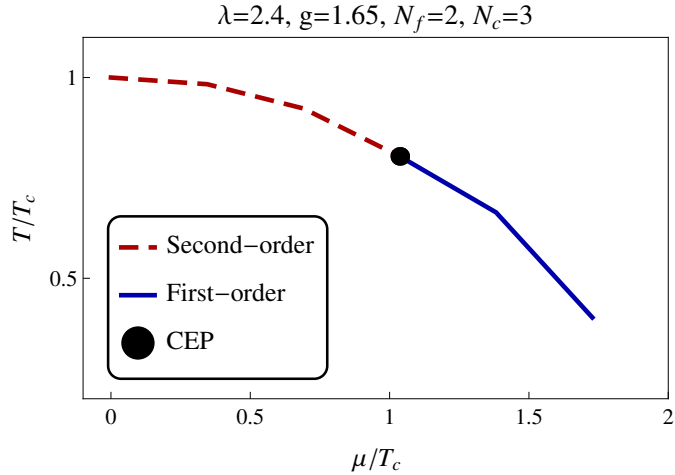}

\caption{(Color on-line) Phase diagram computed from the low-temperature approximation with $T_c/a=1/2$, which corresponds to $m_\sigma\simeq 540$ MeV and $\lambda=2.4$ and $g=1.65$. To set the values of the couplings, we have required that the location of the CEP is in the same region as that obtained in the high-temperature approximation.}
\label{fig6}
\end{figure}

We now turn to study the phase diagram from the low-$T$ approximation. We consider $T_c/a=1/2,$ which corresponds to $m_\sigma\simeq 540$ MeV. Therefore, Eq.~(\ref{relation}) provides a concrete new restriction for the possible values of the couplings. Furthermore, we choose a set of values that place the CEP in the same region as the one we obtained in the high temperature limit, consistent  with mathematical extensions of lattice QCD~\cite{Sharma}. This gives $\lambda=2.4$ and 
$g=1.65$ and Figure~\ref{fig6} shows the phase diagram and the CEP thus found. 
Note that Figs.~\ref{fig5} and~\ref{fig6} describe essentially the same phase diagram.

\begin{figure}[t]
\centering
\includegraphics[width=0.7\textwidth]{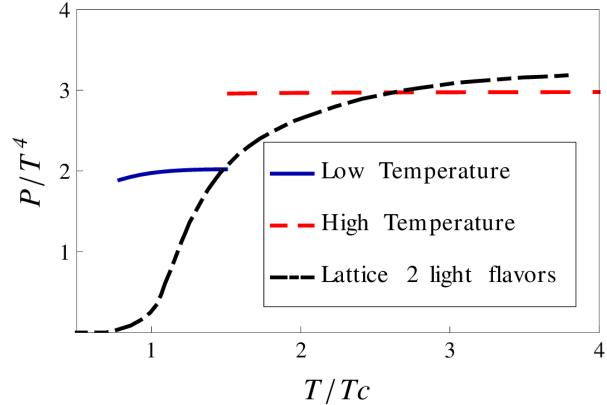}

\caption{(Color on-line) The pressure computed in the model divided by $T^4$ and compared to lattice data for two light flavors. The values of the couplings used in the low and high-temperature approximations are $\lambda_{LT}=2.4$ and $g_{LT}=1.65$, $\lambda_{HT}=0.86$ and $g_{HT}=1.11$, respectively. Note that the model gives an average description of lattice data  for each temperature range, reproducing the jump of the pressure around $T_c$, which is due to the change of the couplings.}
\label{fig7}
\end{figure}

In order to test the consequences of describing the phase diagram with two 
sets of coupling constants we proceed to compute the pressure $P$, also in the low and high-temperature regimes. Recall that the thermodynamical relation between $P$ and $V^{({\mbox{\small{eff}}})}$ is given by
\bea
P = - V^{({\mbox{\small{eff}}})}(v=0).
\label{themrel}
\eea
Figure~\ref{fig7} shows $P/T^4$ computed at $\mu = 0$ in the low- and high-temperature approximations compared to the lattice calculation for two light flavors~\cite{Karsch}. For the low-temperature regime we use the appropriate values of the couplings, namely $\lambda=\lambda_{LT}=2.4$ and $g=g_{LT}=1.65$, whereas for the high-temperature regime we use $\lambda=\lambda_{HT}=0.86$ and $g=g_{HT}=1.11$. Note that the computed pressure provides an average description of lattice data for each temperature range, i.e., the average value of $P/T^4$ at low temperature (up to the largest value of $T/T_c$  that we can reach) is about the average value of lattice data in that temperature range, $P/T^4\simeq 2$. The same is true in the high-temperature description, for which we obtain $P/T^4\simeq 3$. The jump from the low to the high-temperature phases around $T_c$ can be linked to the change in the values of the coupling constants. Furthermore, it can be shown from Eq.~(\ref{Veff-mid}) that when the couplings become smaller the effective potential becomes deeper and thus its negative becomes larger. Thus, the change in the couplings reflects the way the model can effectively incorporate the change in the degrees of freedom, which is generally understood as a change from hadronic to partonic degrees of freedom when going from the low to the high regimes.

\section{Summary and Conclusions}\label{concl}

In this work we have studied the effective QCD phase diagram using the LSMq. We have computed the
finite $T$ and $\mu$-dependent effective potential including the plasma screening effects working
up to the ring diagram order. In the low temperature approximation we fix the couplings by requiring
the CEP location to be the same as the one obtained in the high temperature approximation, 
which we have determined in a previous work~\cite{Ayala1}. In this latter approximation we have 
used the restriction stemming from the condition that relates the couplings to the critical 
temperature for $\mu=0$ and the mass parameter $a$, together with lattice information on $T_{c}$.

The difference between the sets of couplings thus obtained is a measure of change in the 
effective degrees of freedom in each phase.

The phase diagram derived within the high temperature approximation is essentially the same as the one in the low temperature limit. We use this information to calculate the pressure and compare it to lattice data for two light flavors. Though the pressure does not show a total agreement with the lattice results, it provides an average description. We emphasize that the LSMq is an effective QCD theory and that as such its use is limited to provide average values of observables. The change of the pressure with temperature from down below up to above the critical temperature is a signature of QCD that reveals the way the degrees of freedom are activated as the temperature is raised, and this detailed description cannot be captured by an effective model such as
the LSMq. Nevertheless, the change of the pressure curve from the low to the high-temperature descriptions can be attributed to the change of the values of the coupling constants, which reflects the way the model can describe the change of degrees of freedom when going from the hadronic to the quark-gluon phase.

Overall the findings of this work support the idea that the LSMq is an adequate effective analytical tool to describe in average the phase transition in QCD at finite temperature and density. We believe this description can also play a important role in determining the location of the CEP in QCD  in the sense that, as in the case of the LSMq, the infrared properties of the plasma need to be accounted for and furthermore, the identification of the transition curves could be accomplished from knowledge of the behavior of the order parameter for the chiral transition without resorting to studying simultaneously the order parameter for the deconfinement transition.

\section*{Acknowledgments}

The authors acknowledge seminal conversations with R. L. S Farias and G. Krein. Support for this work has been received in part from CONACyT grant number 128534, from UNAM-DGAPA-PAPIIT grant number IN101515 and CIC-UMSNH grant number 4.22.

\end{document}